\documentclass[fleqn,usenatbib]{mnras}

\usepackage{newtxtext,newtxmath}

\usepackage[T1]{fontenc}

\DeclareRobustCommand{\VAN}[3]{#2}
\let\VANthebibliography\thebibliography
\def\thebibliography{\DeclareRobustCommand{\VAN}[3]{##3}\VANthebibliography}

\usepackage{graphicx}

\graphicspath{{./fig/}}
\usepackage{bm}

\newcommand{\EQ}{\begin{equation}}
\newcommand{\EN}{\end{equation}}
\newcommand{\EQA}{\begin{eqnarray}}
\newcommand{\ENA}{\end{eqnarray}}

\newcommand{\Eq}[1]{Eq.~(\ref{#1})}
\newcommand{\Eqs}[2]{Eqs.~(\ref{#1}) and~(\ref{#2})}
\newcommand{\Eqss}[2]{Eqs.~(\ref{#1})--(\ref{#2})}

\newcommand{\App}[1]{Appendix~\ref{#1}}

\newcommand{\Sec}[1]{Sect.~\ref{#1}}
\newcommand{\Secs}[2]{Sects.~\ref{#1} and \ref{#2}}

\newcommand{\Fig}[1]{Fig.~\ref{#1}}

\newcommand{\Tab}[1]{Table~\ref{#1}}

\newcommand{\bra}[1]{\langle #1\rangle}

{}
{}
{}

{}
{}
{}
{}
{}
{}
{}
{}
{}
{}
{}
{}
{}
{}
{}
{}
{}

{}

{}

{}
{}
{}

\newcommand{\BBhat}{\hat{\bm{B}}}

\newcommand{\kk}{\bm{k}}

\newcommand{\BB}{\bm{B}}

\newcommand{\JJ}{\bm{J}}

\newcommand{\AAA}{\bm{A}}

\newcommand{\uu}{\bm{u}}

\newcommand{\ff}{\mbox{\boldmath $f$} {}}

\newcommand{\nab}{{\bm{\nabla}}}

\newcommand{\SSSS}{\mbox{\boldmath ${\sf S}$} {}}

\newcommand{\DD}{{\rm D} {}}

\newcommand{\dd}{{\rm d} {}}

\def\ga{\mathrel{\mathchoice {\vcenter{\offinterlineskip\halign{\hfil
$\displaystyle##$\hfil\cr>\cr\sim\cr}}}
{\vcenter{\offinterlineskip\halign{\hfil$\textstyle##$\hfil\cr>\cr\sim\cr}}}
{\vcenter{\offinterlineskip\halign{\hfil$\scriptstyle##$\hfil\cr>\cr\sim\cr}}}
{\vcenter{\offinterlineskip\halign{\hfil$\scriptscriptstyle##$\hfil\cr>\cr\sim\cr}}}}}
\def\Ma{\mbox{\rm Ma}}

\def\Pm{\mbox{\rm Pr}_{\rm M}}
\def\Rm{\mbox{\rm Re}_{\rm M}}

\def\Rey{\mbox{\rm Re}}

\def\EEM{{\cal E}_{\rm M}}

\def\EK{E_{\rm K}}
\def\EM{E_{\rm M}}
\def\EKM{E_{\rm K/M}}

\def\cs{c_{\rm s}}

\def\kf{k_{\rm f}}

\def\EM{E_{\rm M}}

\def\epsK{\epsilon_{\rm K}}
\def\epsM{\epsilon_{\rm M}}

\def\Brms{B_{\rm rms}}

\def\urms{u_{\rm rms}}

\def\Beq{B_{\rm eq}}

\newcommand{\km}{\,{\rm km}}

\def\apj{ApJ}
\def\apjl{ApJL}

\def\pre{PhRvE}

\def\prl{PhRvL}

\def\araa{ARA\&A}

\def\mnras{MNRAS}

\def\nat{Natur}

\title[Scales in small-scale dynamos]{Dissipative magnetic structures and scales in small-scale dynamos}

\author[A. Brandenburg et al.]{
Axel Brandenburg$^{1,2,3,4}$\thanks{E-mail:brandenb@nordita.org},
Igor Rogachevskii$^{5,1}$, and
Jennifer Schober$^6$
\\
$^1$Nordita, KTH Royal Institute of Technology and Stockholm University, Hannes Alfv\'ens v\"ag 12, SE-10691 Stockholm, Sweden \\
$^2$The Oskar Klein Centre, Department of Astronomy, Stockholm University, AlbaNova, SE-10691 Stockholm, Sweden\\
$^3$McWilliams Center for Cosmology and Department of Physics, Carnegie Mellon University, 5000 Forbes Ave, Pittsburgh, PA 15213, USA\\
$^4$School of Natural Sciences and Medicine, Ilia State University, 3-5 Cholokashvili Avenue, 0194 Tbilisi, Georgia\\
$^5$Department of Mechanical Engineering, Ben-Gurion University of the Negev, P.O. Box 653, Beer-Sheva 84105, Israel\\
$^6$Laboratoire d'Astrophysique, EPFL, CH-1290 Sauverny, Switzerland
}

\date{\today}

\pubyear{2022}

\begin{document}
\label{firstpage}
\pagerange{\pageref{firstpage}--\pageref{lastpage}}
\maketitle

\begin{abstract}
Small-scale dynamos play important roles in modern astrophysics, especially
on Galactic and extragalactic scales.
Owing to dynamo action, purely hydrodynamic Kolmogorov turbulence hardly
exists and is often replaced by hydromagnetic turbulence.
Understanding the size of dissipative magnetic structures is important
in estimating the time scale of Galactic scintillation and other
observational and theoretical aspects of interstellar and intergalactic
small-scale dynamos.
Here we show that, during the kinematic phase of the small-scale
dynamo, the cutoff wavenumber of the magnetic energy spectra scales as
expected for large magnetic Prandtl numbers, but continues in the same
way also for moderately small values -- contrary to what is expected.
For a critical magnetic Prandtl number of about 0.3, the dissipative
and resistive cutoffs are found to occur at the same wavenumber.
In the nonlinearly saturated regime, the critical magnetic Prandtl number
becomes unity.
The cutoff scale now has a shallower scaling with magnetic Prandtl number
below a value of about three, and a steeper one otherwise compared to the kinematic regime.
\end{abstract}


\section{Introduction}

Since the early 1990s, we know that dissipative structures in hydrodynamic
turbulence are vortex tubes \citep{SJO90,VM91}.
Their typical size is of the order of the Kolmogorov length.
In magnetohydrodynamics (MHD), the dissipative structures are magnetic
flux tubes \citep{Nor+92,Bra+96, Mof+94, PP98}.
Their thickness has been estimated to scale with the magnetic Prandtl
number $\Pm=\nu/\eta$, i.e., the ratio of the kinematic viscosity $\nu$
to the magnetic diffusivity $\eta$.
\cite{BPS95}, hereafter BPS, estimated the typical coherence scale
of magnetic field vectors in terms of the gradient matrix
$\nab\BBhat$ of the unit vector $\BBhat=\BB/|\BB|$ of the magnetic
field $\BB$ and found that it scales like $\Pm^{-1/2}$ relative to the Kolmogorov length scale.
The inverse length scale of the magnetic structures can be calculated
as the rms value of $\nab\BBhat$, i.e., $k_B=\bra{|\nab\BBhat|^2}^{1/2}$.
The simulations of BPS were for the case of a convection-driven
dynamo in the presence of rotation and compressibility, but similar
results were later also obtained by \cite{Scheko+04} for a small-scale
dynamo in homogeneous incompressible turbulence for $\Pm\leq1$.
They also emphasized that a steeper dependence on $\Pm$ is expected
for $\Pm\ll1$.

The mechanisms of the small-scale dynamo action are different depending
on the magnetic Prandtl number.
For $\Pm \gg 1$, self-excitation of magnetic fluctuations is caused
by the random stretching of the magnetic field by a smooth velocity
field; see the analytical studies by \cite{Kaz68}, \cite{ZKR90},
\citet[][hereafter KA]{Kulsrud+Anderson92}, and \cite{SSF12}.
For $\Pm \ll 1$, the small-scale dynamo is driven by velocity
fluctuations at the resistive scale, which is located in the
inertial range \citep{Kaz68,RK97,BC04,AH07,KR12,AMV19}.
In particular, KA found that, for large values of
$\Pm$, the magnetic energy spectrum is expected to be of the form
\begin{equation}
\EM(k,t)\propto e^{2\gamma t} k^{3/2} K_0\left(k/k_\eta^{\rm KA} \right),
\label{EMk0}
\end{equation}
where $K_0$ is the Macdonald function of order zero 
(or the modified Bessel function of the second kind),
and $k_\eta^{\rm KA}$ is
\begin{equation}
k_\eta^{\rm KA}=(4\gamma/15\eta)^{1/2},
\label{ketaKA}
\end{equation}
where $\gamma$ is the growth rate of the magnetic field.\footnote{Note
that the symbol $\gamma$ used in KA is 3/8th of the growth rate,
while the $\gamma$ used here is the actual growth rate.}
This provides another very different method for calculating a relevant
wavenumber characterizing the scale of structures than $k_B$.

The question of characteristic length scales in a small-scale dynamo
continued attracting attention and has been investigated in more detail
by \cite{CR09} with applications to the intergalactic medium.
Much of this work concerns the saturated phase of the dynamo, but
\Eq{EMk0} is then not applicable.
More recently, \cite{Kriel+22} confirmed the $\Pm^{-1/2}$ scaling
for $1\leq\Pm\leq260$ for the kinematic phase of the dynamo.
The small-scale properties of interstellar turbulence can be
assessed through interstellar scintillation measurements of pulsars
\citep{Cordes+85, Rickett90, Armstrong+95, Bhat+04, Scalo+Elmegreen04}.
A particular difficulty is to explain what is known as extreme scattering
events (ESEs), which would require unrealistically large pressures if
the scattering structures were spherical \citep{Clegg+98}.
This favors the presence of sheet- or tube-like structures that could
explain ESEs of those structures are oriented along the line-of-sight
\citep{Pen+King12, Bannister+16}.
Scintillation measurements suggest that the dissipative structures
of MHD turbulence are sheet-like with an inner scale down to
$300\km$ \citep{Bhat+04}.
However, more detailed measurements would be needed to determine the
precise nature of the smallest dissipative structures \citep{Xu+Zhang17}.

The goal of the present paper is to compare the relations between 
different length scales in small-scale dynamos.
We mainly focus here on the kinematic growth phase of the dynamo, but
we also consider some nonlinear models in
\Secs{GrowthPhase}{NonlinPrandtl}.
In addition to the values of $k_\eta^{\rm KA}$ and $k_B$ discussed above,
we also determine a wavenumber $k_\eta$ that describes the resistive
cutoff of the spectrum, and is analogous to the wavenumber $k_\nu$
based on the Kolmogorov (viscous) scale.
\cite{Kriel+22} used a similar prescription, but did not compare with
other magnetic scales.
Note that, contrary to $k_\eta^{\rm KA}$, $k_\eta$ is not calculated
from the dynamo growth rate.
Following earlier work \citep{Bra+18}, we consider weakly compressible
turbulence with an isothermal equation of state and a constant sound
speed $\cs$, where the pressure is proportional to the density $\rho$,
i.e., $p=\rho\cs^2$.

\section{The model}

\subsection{Basic equations}

In this work, we are primarily interested in weak magnetic fields and
ignore therefore the Lorentz force in most simulations.
The magnetic field is given as $\BB=\nab\times\AAA$, where $\AAA$
is the magnetic vector potential.
We thus solve the evolution equations for the magnetic vector potential
$\AAA$, the velocity $\uu$, and the logarithmic density $\ln\rho$ in
the form
\begin{equation}
\frac{\partial\AAA}{\partial t}=\uu\times\BB+\eta\nabla^2\AAA,
\label{dAdt}
\end{equation}
\begin{equation}
\frac{\DD\uu}{\DD t}=\ff-\cs^2\nab\ln\rho+
\frac{1}{\rho}\nab\cdot(2\rho\nu\SSSS),
\label{dudt}
\end{equation}
\begin{equation}
\frac{\DD\ln\rho}{\DD t}=-\nab\cdot\uu,
\label{dlnrhodt}
\end{equation}
where $\DD/\DD t=\partial/\partial t+\uu\cdot\nab$
is the advective derivative,
$\ff$ is a nonhelical forcing function consisting
of plane waves with wavevector $\kk$, and ${\sf S}_{ij}=
(\partial_i u_j+\partial_j u_i)/2-\delta_{ij}\nab\cdot\uu/3$
are the components of the rate-of-strain tensor $\SSSS$.
For the forcing, we select a $\kk$ vector at each time step randomly
from a finite shell around $\kf/k_1=1.5$ with $1\leq|\kk|/k_1<2$.
The components of $\kk$ are taken to be integer multiples of $k_1\equiv2\pi/L$,
where $L$ is the side length of our Cartesian domain of volume $L^3$.
This forcing function has been used in many earlier papers
\citep[e.g.][]{HBD04}.
We solve \Eqss{dAdt}{dlnrhodt} using the {\sc Pencil Code} \citep{JOSS}.

\subsection{Spectra and characteristic parameters}
\label{Spectra}

We normalize our kinetic and magnetic energy spectra
such that $\int\EK(k)\,\dd k=\bra{\uu^2}/2$ and
$\int\EM(k)\,\dd k=\bra{\BB^2}/2\mu_0\rho_0\equiv\EEM$,
respectively, where $\rho_0$ is the mean density.
Here, angle brackets without subscript denote volume averages.
We always present time-averaged spectra.
Since $\EM(k,t)$ increases exponentially at the rate $2\gamma$,
where $\gamma$ is the growth rate of the magnetic field, we average
the compensated spectra, $\bra{e^{-2\gamma t}\EM(k,t)}_{\Delta t}$,
over a suitable time interval $\Delta t$ where the function
$e^{-2\gamma t}\EM(k,t)$ is statistically stationary; see also \cite{SB14}.
Our averaged magnetic energy spectra are normalized by $\EEM$, so that
their integral is unity.

Our governing parameters are the Mach number, and the fluid and magnetic
Reynolds numbers, defined here as
\begin{equation}
\Ma=\urms/\cs,\quad
\Rey=\urms/\nu\kf,\quad
\Rm=\urms/\eta\kf,
\end{equation}
respectively, where $\urms$ is the time-averaged rms velocity.
Thus, the magnetic Prandtl number is $\Pm=\Rm/\Rey$.
The value of $\gamma$ is computed as the average of $\dd\ln\Brms/\dd t$
during the exponential growth phase.
We also give the kinetic dissipation wavenumber
\begin{equation}
k_\nu=\left(\epsK/\rho_0\nu^3\right)^{1/4},
\label{knudef}
\end{equation}
where $\epsK=\bra{2\rho\nu\SSSS^2}_{\Delta t}$ 
is the time-averaged kinetic energy
dissipation rate.
It obeys the expected $\Rey^{3/4}$ scaling, here with
$k_\nu/\kf\approx0.48\,\Rey^{3/4}$; see \App{knuScaling}.

In fluid dynamics, to avoid discussions about different definitions
of the Reynolds number, one commonly quotes the Taylor microscale
Reynolds number \citep{TL72}, which is universally defined as
$\Rey_\lambda=u'\lambda_{\rm Tay}/\nu$.
Here, $u'=\urms/\sqrt{3}$ is the one-dimensional rms velocity and
$\lambda_{\rm Tay}=\sqrt{15\nu\rho_0/\epsK}\,u'$ is the Taylor
microscale.\footnote{We correct herewith a typo in \cite{Haugen+22},
where the $u'$ factor in $\lambda_{\rm Tay}$ was dropped in their
definition, but it was included in their calculations.}
The values of $\Rey_\lambda$ are given in \Tab{Tsum}.
They are expected to be proportional to $\Rey^{1/2}$, but the
actual scaling is slightly steeper; see the Supplemental Material
\citep{DATA}.

A tilde on the growth rate denotes normalization with the 
turnover rate and tildes on various wavenumbers 
denote normalization with respect to $k_1$, i.e.,
\begin{equation}
\tilde{\gamma}=\gamma\tau,\quad
\tilde{k}_\nu=k_\nu/k_1,\quad
\tilde{k}_{\rm f}=\kf/k_1,\quad
\mbox{etc},
\end{equation}
where $\tau=1/\urms\kf$ is the turnover time.
These parameters are listed in \Tab{Tsum} for our runs.
For Runs~A--K, we used a resolution of $512^3$ mesh points,
while we used $1024^3$ mesh points for Runs~L and M,
and $2048^3$ mesh points for Run~M'.
The value of $\epsK$ in units of $\rho_0 k_1\cs^3$ is obtained
from the table entries as $\epsK/\rho_0 k_1\cs^3=\tilde{k}_\nu^4
(\Ma/\Rey\,\tilde{k}_{\rm f})^3$.
The calculation of the values of $\tilde{k}_\eta$ is discussed below.
Error bars are computed from time series as the largest departure
of any one third compared to the total.

In some cases, we examine the effects of nonlinear saturation.
We then include the Lorentz force and replace \Eq{dudt} by
\begin{equation}
\frac{\DD\uu}{\DD t}=\ff-\cs^2\nab\ln\rho+
\frac{1}{\rho}\Big[\nab\cdot(2\rho\nu\SSSS)+\JJ\times\BB\Big].
\label{dudtLor}
\end{equation}
Once the Lorentz force is included, the magnetic field is expected to
saturate near the equipartition magnetic field strength,
$\Beq=\sqrt{\mu_0\rho_0}\urms$.

\begin{table*}
\centering
\caption{
Summary of the kinematic simulations presented in this paper.
}\label{Tsum}
\begin{tabular}{ccrrrccccccrr} 
\hline
$\!\!$Run$\!\!$ & $\Ma$ & $\Rey_\lambda$ & $\Rey$ & $\Rm$ & $\Pm$ & $\tilde{\gamma}$ & $\tilde{k}_\nu$ & $\tilde{k}_B$ &
$\tilde{k}_\eta^{\rm KA}$ & $\tilde{k}_\eta$ & $\Delta t/\tau$ & $N$ \\
\hline
A & 0.096 &  13 &   12 & 1240 & 100    &$ 0.076\pm0.014 $&$  5.9\pm0.1$&$ 127\pm2 $&$7.7\pm0.7$  & $ 109\pm4 $& 31 & 512 \\
B & 0.113 &  30 &   36 & 1460 &  40    &$ 0.090\pm0.006 $&$ 11.7\pm0.1$&$ 128\pm4 $&$9.1\pm0.3$  & $ 129\pm4 $& 62 & 512 \\
C & 0.120 &  50 &   78 & 1560 &  20    &$ 0.110\pm0.002 $&$ 19.7\pm0.2$&$ 139\pm3 $&$10.4\pm0.2$ & $ 156\pm6 $& 87 & 512 \\
D & 0.127 &  70 &  165 & 1650 &  10    &$ 0.135\pm0.006 $&$ 34.1\pm0.2$&$ 156\pm3 $&$11.8\pm0.2$ & $ 187\pm8 $& 98 & 512 \\
E & 0.130 & 120 &  420 & 1680 &   4    &$ 0.159\pm0.007 $&$   68\pm2  $&$ 185\pm3 $&$13.0\pm0.3$ & $ 248\pm10$& 75 & 512 \\
F & 0.128 & 170 &  830 & 1660 &   2    &$ 0.172\pm0.014 $&$  113\pm5  $&$ 209\pm6 $&$13.4\pm0.5$ & $ 293\pm15$& 62 & 512 \\
G & 0.129 & 250 & 1670 & 1670 &   1    &$ 0.157\pm0.016 $&$  185\pm7  $&$ 237\pm4 $&$12.9\pm0.6$ & $ 358\pm15$& 43 & 512 \\
G'& 0.131 & 250 & 1700 & 1700 &   1    &$ 0.144\pm0.020 $&$  188\pm10 $&  ...      &$12.5\pm0.6$ & $ 358\pm15$& 54 &1024 \\
H & 0.132 & 260 & 1710 &  850 &   0.5  &$ 0.079\pm0.006 $&$  187\pm6  $&$ 147\pm4 $&$6.5\pm0.3$  & $ 260\pm15$&101 & 512 \\
I & 0.134 & 260 & 1740 &  580 &   0.33 &$ 0.042\pm0.010 $&$  189\pm5  $&$ 114\pm3 $&$3.9\pm0.5$  & $ 216\pm15$& 78 & 512 \\
J & 0.130 & 260 & 1680 &  420 &   0.25 &$ 0.029\pm0.001 $&$  185\pm3  $&$  92\pm1 $&$2.8\pm0.3$  & $ 189\pm20$&712 & 512 \\
K & 0.130 & 250 & 1680 &  340 &   0.20 &$ 0.019\pm0.004 $&$  186\pm5  $&$  82\pm4 $&$2.0\pm0.2$  & $ 168\pm20$& 99 & 512 \\
L & 0.132 & 420 & 4270 &  427 &   0.10 &$ 0.020\pm0.003 $&$  368\pm10 $&$ 107\pm2 $&$2.3\pm0.3$  & $ 249\pm18$&193 &1024 \\
M & 0.132 & 650 & 8300 &  430 &   0.05 &$ 0.013\pm0.008 $&$  575\pm17 $&$ 103\pm4 $&$1.8\pm0.4$  & $ 332\pm15$& 61 &1024 \\
M'& 0.134 & 620 & 8700 &  430 &   0.05 &$ 0.016\pm0.011 $&$  610\pm12 $&$ 103\pm4 $&$1.8\pm0.4$  & $ 332\pm15$&  6 &2048 \\
\hline
\end{tabular}
\end{table*}

\section{Results}

\subsection{Growth phase of the dynamo}
\label{GrowthPhase}

In most of this work, we analyze kinematic dynamo action, i.e., the
Lorentz force is weak and can be neglected.
This means that the magnetic field of a supercritical dynamo grows
exponentially beyond any bound.

\begin{figure}\begin{center}
\includegraphics[width=\columnwidth]{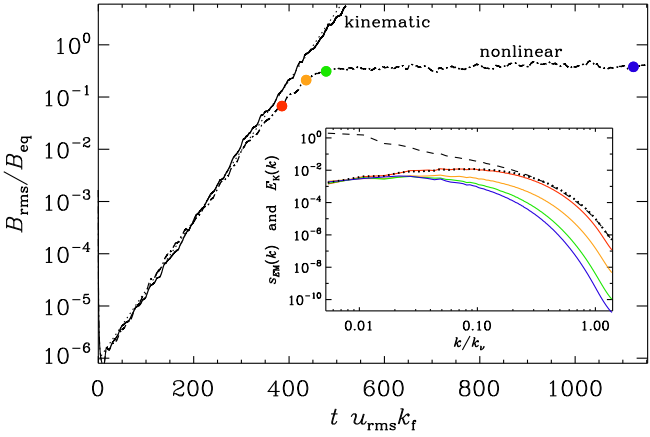}
\end{center}\caption[]{
Nonlinear saturation (Run~j of \Sec{NonlinPrandtl} below) compared
with the kinematic evolution for Run~J (see \Tab{Tsum}).
The red, orange, green, and blue dots mark the times when
$\Brms/\Beq\approx0.07$, 0.2, 0.3, and 0.4.
The inset shows scaled magnetic energy spectra, $s_E\EM(k)$,
where $s_E=16$, 0.45, 0.11, and 0.06, so as to make the spectra
overlap near the smallest wavenumber.
The dashed and dotted lines give the time-averaged spectra
$\EK(k)$ and $\EM(k)$, respectively in the kinematic regime
without the Lorentz force.
}\label{pcomp_kin}\end{figure}

To quantify the point until when the Lorentz force can indeed be
neglected, we present in this section simulations with the Lorentz force
included; see \Eq{dudtLor}.
We then expect the magnetic field to saturate near $\Beq$.
In \Fig{pcomp_kin}, we show the evolution of $\Brms/\Beq$ for cases
with and without Lorentz force included.
We also mark four particular times for which we also show the magnetic
energy spectra in the nonlinear regime.
We see that, when $\Brms/\Beq\approx0.05$, the magnetic energy spectrum
(red line) is still close to the time-averaged kinematic spectrum
(dotted line).
At the time when $\Brms/\Beq\approx0.2$, 
we begin to see 
clear departures from the kinematic spectrum $\EM(k)$.
To see this more clearly, we have scaled the amplitude of the spectra
such that they agree with the kinematic one (dotted line) near the
smallest wavenumber.
Finally, when $\Brms/\Beq\approx0.3$, a slow phase of nonlinear saturation
commences where the value of $\Brms/\Beq$ hardly changes, but the spectrum
still changes in such a way that its peak moves into the inertial range.
This is an important difference to the kinematic stage and was first
report by \cite{HBD03}.
The final value of $\Brms/\Beq$ is about 0.4.

\subsection{Scalings of the KA and flux tube wavenumbers}

Looking at \Tab{Tsum}, it is clear\footnote{Note that
$\gamma/\eta k_1^2=\tilde{\gamma}\Rm\,\tilde{k}_{\rm f}^2$
is related to values given in \Tab{Tsum}.}
that the inverse flux tube thickness
$\tilde{k}_B$ does not change monotonically with $\Pm$.
The same is also true for $\tilde{k}_\eta^{\rm KA}$.
This is mostly because $\Rm$ was not kept constant for all runs.
For $\Pm\ge1$, however, $\Rm$ varied only little and was in the
range from 1200 to 1700.
In that range, $\tilde{k}_B$ showed a steady increase with $\Pm$.
For smaller $\Pm$, we decrease $\Rm$ so that $\Rey$ did not become
too large.
For Runs~L and M, we used a resolution of $1024^3$ and were thus able
to increase $\Rey$, which led to a slight increase of $\tilde{k}_B$.
For Run~M', we used $2048^3$ mesh points and find results comparable
to those of Run~M, except for the larger statistical error.
In most of the plots, we normalize the characteristic wavenumbers
by $k_\nu$, which resulted in a monotonic increase of the ratios
$k_B/k_\nu$ and $k_\eta^{\rm KA}/k_\nu$.

\begin{figure}\begin{center}
\includegraphics[width=\columnwidth]{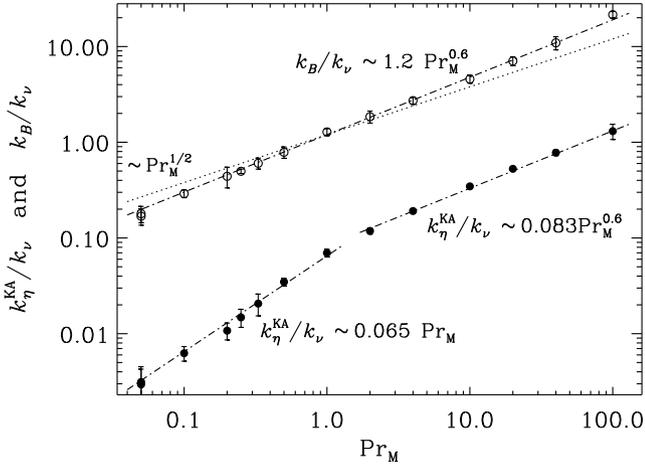}
\end{center}\caption[]{
Dependence of $k_\eta^{\rm KA}/k_\nu$ (closed symbols) and
$k_B/k_\nu$ (open symbols) on $\Pm$.
The dotted line shows the $\Pm^{1/2}$ scaling for comparison.
}\label{ptable_keta}\end{figure}

In \Fig{ptable_keta}, we plot $k_\eta^{\rm KA}/k_\nu$ and $k_B/k_\nu$
versus $\Pm$.
Both show a $\Pm^{0.6}$ scaling for $\Pm\geq 2$, but they have a linear
dependence for $\Pm < 1$.
Thus, the expected $\Pm^{1/2}$ scaling is only approximately confirmed.

\subsection{Resistive cutoff wavenumbers}

Important characteristics of MHD turbulence are the kinetic and
magnetic energy spectra.
Focussing on the viscous and resistive dissipation subranges, it
makes sense to normalize $k$ by $k_\nu$, as discussed above.
We recall that the quantity $k_\nu$ is usually defined as
in \Eq{knudef}, i.e., without any prefactors.
The point when the spectrum drops significantly is typically at
$k/k_\nu\approx0.1$ rather than at unity, as one might have expected.
This should be kept in mind when discussing values of
cutoff wavenumbers in other definitions.
We return to this at the end of the paper.

\begin{figure}\begin{center}
\includegraphics[width=\columnwidth]{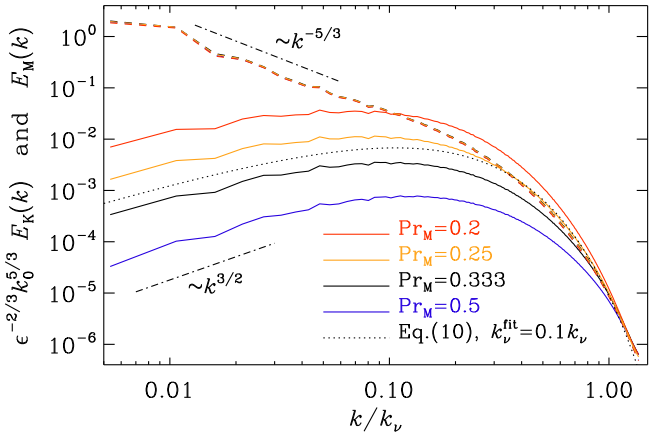}
\end{center}\caption[]{
Magnetic energy spectra (solid lines) for $\Pm=1/5$ (Run~K, red),
1/4 (Run~J, orange), 1/3 (Run~I, black), and 1/2 (Run~H, blue) along
with the corresponding kinetic energy spectra (dashed lines).
}\label{pspec_comp_PrMcrit}\end{figure}

The functional forms of $\EM(k)$ and $\EK(k)$ are rather different at
small values of $k$, but near the viscous cutoff wavenumber they are
more similar to each other.
In \Fig{pspec_comp_PrMcrit}, we compare $\EK(k)$ and $\EM(k)$ for a
few values of $\Pm$.
We clearly recognize the $\EK(k)\propto k^{-5/3}$ Kolmogorov scaling and
the $\EM(k)\propto k^{3/2}$ spectrum of the small-scale dynamo
\citep{Kaz68}; see also KA.
For different values of $\Pm$, however, the slopes of $\EM(k)$ are
quite different near the resistive cutoff wavenumber: steeper for
small values of $\Pm$ and shallower for larger values of $\Pm$.
For $\Pm=1/4=0.25$, the shapes of $\EM(k)$ and $\EK(k)$ are most
similar to each other at large $k$, although $\EM(k)$ is just slightly
too steep, while for $\Pm\ge1/3$, it is already clearly too shallow.
Thus, we expect there to be a critical value, $\Pm^{\rm crit}$ of about
0.3, where $\EM(k)$ and $\EK(k)$ are most similar to each other near
the cutoff wavenumber.

\begin{figure}\begin{center}
\includegraphics[width=\columnwidth]{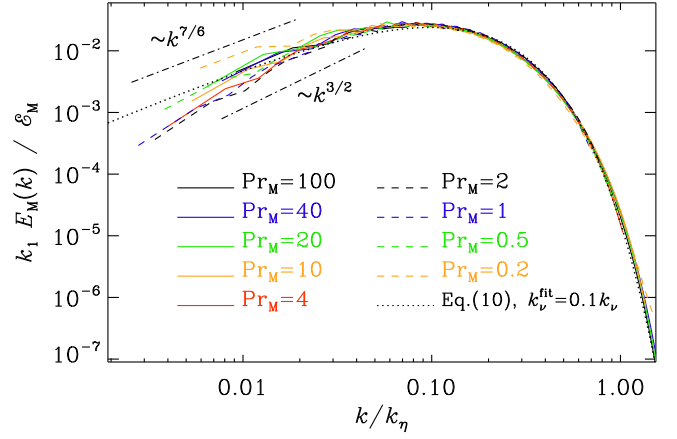}
\end{center}\caption[]{
Magnetic energy spectra collapsed on top of each other by choosing
suitable values of $k_\eta$ for each value of $\Pm$.
The dotted line shows \Eq{DottedLineFit} with
$k_\eta^{\rm fit}=0.13\,k_\eta$.
}\label{pspec_comp2}\end{figure}

\begin{figure}\begin{center}
\includegraphics[width=\columnwidth]{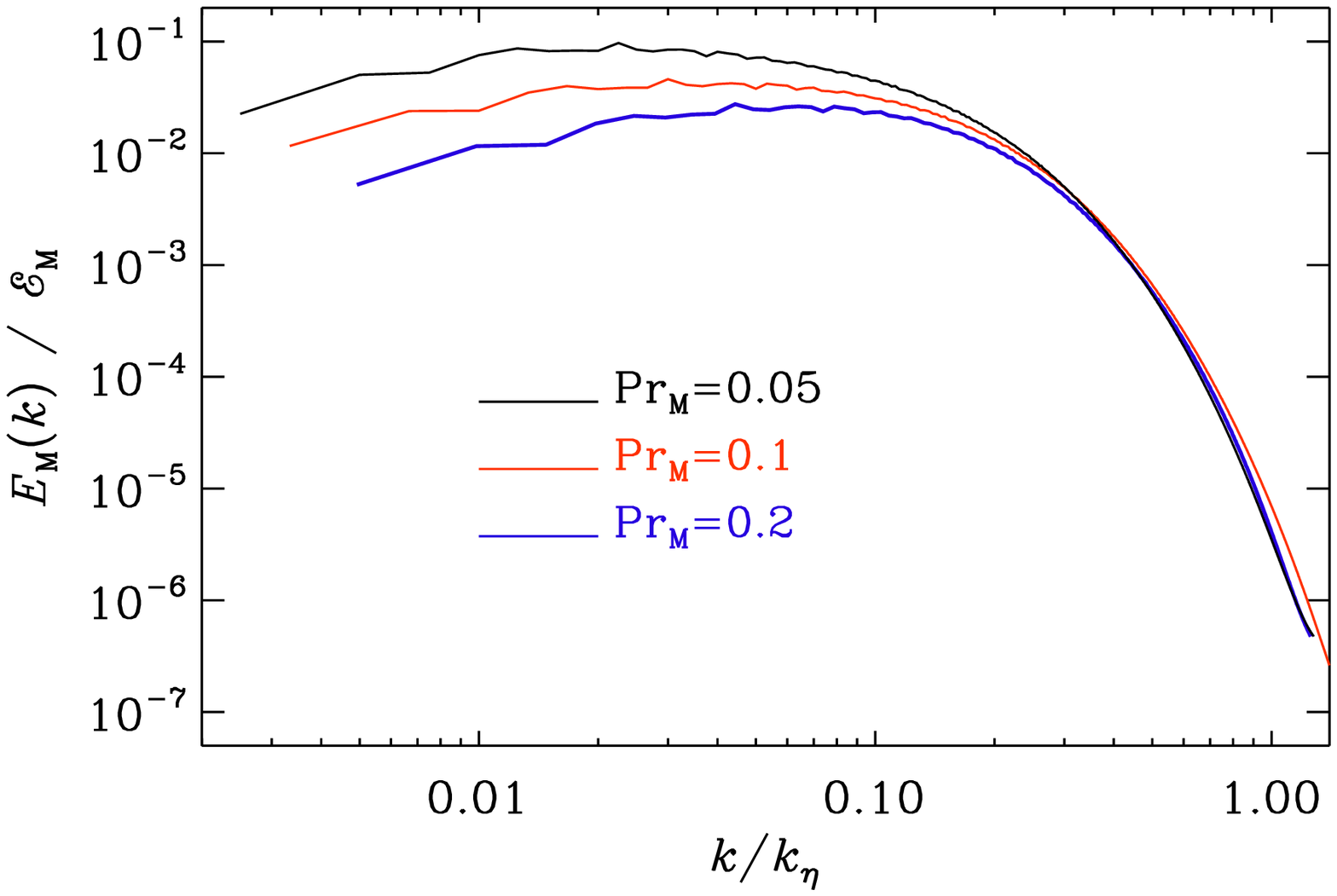}
\end{center}\caption[]{
Similar to \Fig{pspec_comp2}, but for $\Pm=0.05$ (black line),
0.1 (red line), and 0.2 (blue line).
}\label{pspec_comp2_loPm}\end{figure}

The spectral behavior near the resistive cutoff can be compared with
\Eq{EMk0} using an empirical fit parameter through
\begin{equation}
k_1\EM(k)/\EEM=A_0 \, \left(k/k_\eta^{\rm fit}\right)^{3/2}
K_0\left(k/k_\eta^{\rm fit}\right),
\label{DottedLineFit}
\end{equation}
where $k_\eta^{\rm fit}$ is now treated as an adjustable parameter.
In \Fig{pspec_comp_PrMcrit}, we have already compared with
\Eq{DottedLineFit}, although the match is not very good.
This is mostly because the model applies to large values
of $\Pm$, and then the fit improves, as we will see below.

By choosing suitable values of $k_\eta$ for $\Pm\neq\Pm^{\rm crit}$, we
can now try to collapse the curves $\EM(k/k_\eta)$ on top of each other.
This is done in \Fig{pspec_comp2}, where we use Run~I with $\Pm=0.33$
as references run, because this value is close to $\Pm^{\rm crit}$.
The collapse is good near and above the peak of the spectra, but there
are departures for small values of $k$, where the spectra become shallower
than the classical Kazantsev slope for smaller values of $\Pm$.
In the opposite limit of $\Pm\ll1$, the spectral slope may be smaller.
For $\Pm=0.1$, a $k^{7/6}$ scaling was previously discussed by \cite{SB14}
and confirmed by \cite{Bra+18}.
For $\Pm\leq0.2$, the quality of the collapse onto \Eq{EMk0} becomes
rather poor, which is why we plot the results for smaller values
separately; see \Fig{pspec_comp2_loPm}.

\begin{figure}\begin{center}
\includegraphics[width=\columnwidth]{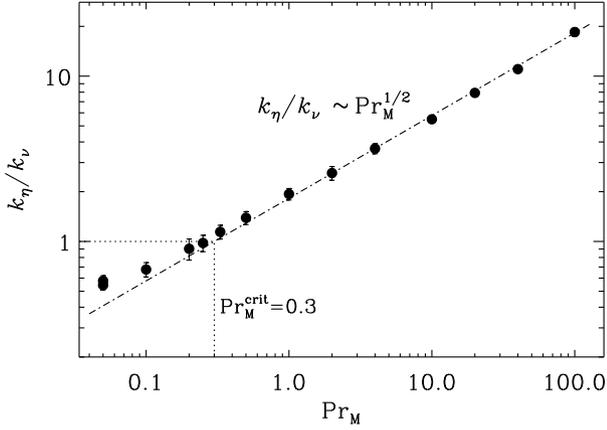}
\end{center}\caption[]{
Dependence of $k_\eta/k_\nu$ (closed symbols) on $\Pm$.
}\label{ptable_keta2}\end{figure}

The collapse for each value of $\Pm$ results in a value of $k_\eta$,
which we have listed in \Tab{Tsum}.
A plot of $k_\eta/k_\nu$ versus $\Pm$ is given in \Fig{ptable_keta2}.
We see that the ratio $k_\eta/k_\nu$ does obey the expected $\Pm^{1/2}$
scaling rather well.
In this figure, we have also highlighted the value of
$\Pm=\Pm^{\rm crit}\approx0.3$ where $k_\eta/k_\nu=1$, so
\begin{equation}
k_\eta/k_\nu=\left(\Pm/\Pm^{\rm crit}\right)^{1/2}.
\label{ketaknu}
\end{equation}
For $\Pm\ll1$, a steeper scaling is numerically obtained at very high
resolution simulations \citep{W22}, 
but in our simulations, such a trend cannot yet be seen
for $\Pm\ge0.05$.

\begin{figure}\begin{center}
\includegraphics[width=\columnwidth]{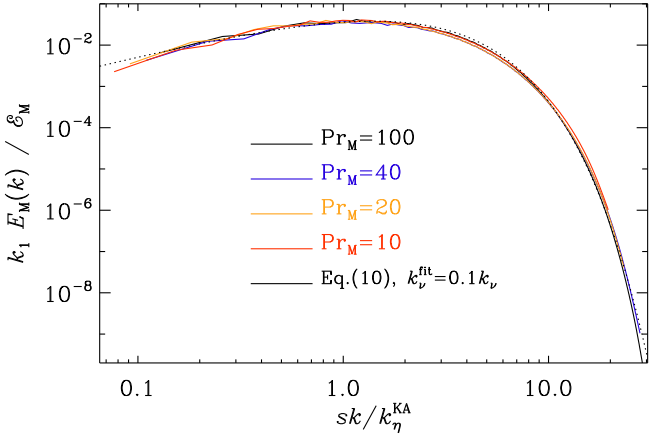}
\end{center}\caption[]{
Magnetic energy spectra versus $k/k_\eta^{\rm KA}$ for $\Pm=100$ (black
solid line) and collapsed on top of it the result for $\Pm=40$ (blue
line), as well as $\Pm=20$ (orange line, having scaled $k_\eta^{\rm KA}$
by a factor 1.05), and $\Pm=10$ (red line, having scaled $k_\eta^{\rm KA}$
by a factor 1.1).
The dotted line shows \Eq{DottedLineFit} with
$k_\eta^{\rm fit}=1.32\,sk_\eta^{\rm KA}$.
}\label{pkaz_all}\end{figure}

\subsection{Comparison with the Kazantsev cutoff wavenumber}

We have already discussed the differences in the $\Pm$ scaling between
the measured $k_\eta$ and the theoretically expected $k_\eta^{\rm KA}$
from the work of KA based on the numerically determined growth rate.
Here, however, the scales are rather different in an absolute sense.
This is primarily caused by the large departure between the values of
$k_\eta^{\rm KA}$ and the location where the magnetic energy spectrum
begins to drop rapidly.
The apparent discrepancy can be alleviated by redefining $k_\nu$ such
that the drop occurs closer to unity.
Thus, there is otherwise no physical significance in the difference of
the absolute wavenumbers.

To clarify this point, we now plot $k_1\EM(k)/\EEM$ versus
$k/k_\eta^{\rm KA}$ for $\Pm=100$ and 40.
For smaller values of $\Pm$, we have scaled $k_\eta^{\rm KA}$ by a
factor $s=1.05$ for $\Pm=20$ and by a factor $s=1.1$ for $\Pm=10$;
see \Fig{pkaz_all}.
Those coefficients are also listed in \Tab{Tsum2}.
The result for $\Pm=4$ is not plotted because of a poor collapse at small $k$.
Here, the adjustment factor is 1.3, as listed in \Tab{Tsum2}.
This lack of collapse for $\Pm\leq4$ illustrates that only for large
values of $\Pm$, the Kazantsev model reproduces the numerical data
related to $k_\eta^{\rm KA}$ sufficiently well.

\begin{table}
\centering
\caption{
Values of $k_\eta^{\rm KA}$ and adjustment factors to the KA values for
$\Pm\leq20$.
}\label{Tsum2}
\begin{tabular}{c|ccccc} 
\hline
$\Pm$                           & 100  & 40    & 20    & 10   &  4   \\
\hline
$\tilde{k}_\eta^{\rm KA}$       &  7.7 &  9.1  & 10.4  & 11.9 & 13.0 \\
$s$                             &   1  &  1.00 &  1.05 &  1.1 &  1.3 \\
$1.3\,s\tilde{k}_\eta^{\rm KA}$ & 10.0 & 11.8  & 14.1  & 16.9 & 21.9 \\
$k_\eta^{\rm fit}/k_\nu$        & 2.87 & 1.44  & 0.56  & 0.38 & 0.25 \\
\hline
\end{tabular}
\end{table}

In absolute terms, the value of $k_\eta^{\rm KA}$ given by \Eq{ketaKA}
underestimates the position of the peak by a factor of about $1.3=1/0.77$.
This factor was obtained empirically by having overplotted in
\Fig{pkaz_all} the graph of \Eq{DottedLineFit} with
\begin{equation}
k_\eta^{\rm fit}
\approx 0.093\,k_\eta
\approx 1.32\,sk_\eta^{\rm KA}.
\label{Params}
\end{equation}
The agreement with the numerical solutions is generally good, but
deteriorates for $\Pm<10$, especially for small $k$, where the
simulation results predict less power than the Kazantsev model.
On the other hand, the discrepancy with the estimate of $k_\eta^{\rm KA}$
decreases owing to the increase of the correction factor $s$,
which is caused by the $\Pm^{0.6}$ scaling found in \Fig{ptable_keta2} for
$\Pm\ga0.2$, instead of the expected $\Pm^{1/2}$ scaling of \Eq{ketaknu}.
This is illustrated in the third line of \Tab{Tsum2}, where we have
listed the values of $1.3\,s\tilde{k}_\eta^{\rm KA}$.
As before, the tilde denotes normalization by $k_1$.
Finally, we also list in \Tab{Tsum2} the ratios $k_\eta^{\rm fit}/k_\nu$.

\subsection{Different viscous cutoff wavenumbers}
\label{Redefining}

The absolute scale of characteristic and cutoff wavenumbers is a matter
of convention.
The value of $k_\nu$, as defined in \Eq{knudef}, plays an important role
in that it is needed to collapse the kinetic energy spectra on top of
each other; see \App{KineticCutoff}.
For $k_\eta$, one could determine empirically the 
effective wavenumber $k_\eta^{\rm KA}$
in \Eqs{EMk0}{DottedLineFit}, as we have done.
This value turned out to be 1.3 times smaller than that
proposed by KA.
One would then define $1.3\,k_\eta^{\rm KA}$ as a new resistive cutoff wavenumber.
Given that the 1/2 scaling in \Eq{ketaknu} is well obeyed, one could even
redefine $k_\nu$ correspondingly.
Looking at \Tab{Tsum}, we see that for $\Pm=\Pm^{\rm crit}$, we have
$k_\nu/k_1\approx200$.
Furthermore, we see that $1.3\,sk_\eta^{\rm KA}=10$.
Thus, since $10/200=0.05$, we could define a magnetically motivated
value as $k_\nu^{\rm mag}=0.05\,k_\nu$.
The motivation for defining $k_\nu^{\rm mag}$ in terms of the
magnetic energy spectrum was because $\EM(k)$ had a well defined peak,
which is not the case for $\EK(k)$.
However, one could compare with
\begin{equation}
\EK(k)\propto k^{-5/3}\exp(-k/k_\nu^{\rm fit}).
\label{EKfitold}
\end{equation}
This is the approach chosen by \cite{Kriel+22}, who found
$k_\nu^{\rm fit}\approx0.025\,\kf\,\Rey_{\rm KBSF}^{3/4}
\approx0.1\,\kf\,\Rey^{3/4}$, where $\Rey_{\rm KBSF}=2\pi\Rey$
is the Reynolds number based on the characteristic length scale
rather than the characteristic wavenumber $\kf$.
Here, we find
\begin{equation}
k_\nu^{\rm fit}\approx0.5\,k_\nu\approx0.24\,\kf\,\Rey^{3/4};
\label{knufit}
\end{equation}
see \App{knuScaling}.
This is about twice as large as their value.

A problem with \Eq{EKfitold} is that it lacks a description
of the bottleneck.
\cite{SJ93} showed that experimental data can best be fit with an
additional $k^{-1}$ piece, while \cite{Qian84} proposed a formula
based on a closure model of the form
\begin{equation}
\EK(k)\propto k^{-5/3}\left[1+\big(k/k_\nu^{\rm bot}\big)^{n_{\rm bot}}\right]
\exp\left[-\big(k/k_\nu^{\rm dis}\big)^{n_{\rm dis}}\right],
\label{EKfit}
\end{equation}
with adjustable coefficients $k_\nu^{\rm bot}$ and $k_\nu^{\rm dis}$,
and exponents $n_{\rm bot}=2/3$ and $n_{\rm dis}=4/3$, which implies a
$k^{-1}$ scaling of the bottleneck.
His formula was also confirmed by \cite{Dobler+03} using the {\sc Pencil Code}.
A better fit is shown in \App{KineticCutoff}, were $n_{\rm bot}=1.8$
and $n_{\rm dis}=0.86$ with $k_\nu^{\rm bot}\approx0.056\,k_\nu$ and
$k_\nu^{\rm dis}\approx0.073\,k_\nu$ were found, which would motivate
another definition; see \Tab{Tsum3} for a summary of the different
cutoff wavenumbers discussed in this paper.

\begin{table}
\centering
\caption{
Summary of the different characteristic wavenumbers used in the paper.
}\label{Tsum3}
\begin{tabular}{l|ll} 
\hline
Quantity           & Definition & Equation \\
\hline
$k_\nu$            & $(\epsK/\rho_0\nu^3)^{1/4}$   & \Eq{knudef} \\
$k_\eta^{\rm NL}$  & $(\epsM/\rho_0\eta^3)^{1/4}$  & \Eq{ketadef} \\
$k_\eta^{\rm KA}$  & $(4\gamma/15\eta)^{1/2}$      & \Eq{ketaKA} \\
$k_\eta^{\rm fit}$ & collapse in \Fig{pspec_comp2} & \Eq{DottedLineFit} \\
$k_\nu^{\rm fit}$  & $\EK(k)\propto k^{-5/3}\exp(-k/k_\nu^{\rm fit})$  & \Eq{EKfitold} \\
$k_\eta^{\rm dis}$ & fit in \Fig{pspec_comp2_kin}(b) & \Eq{EKfit} \\
$k_{\nu/\eta}^{\rm NLfit}$ & $\EKM(k)\propto k^{-1}\exp(-k/k_{\nu/\eta}^{\rm NLfit})$ & \Eq{EKfitNL} \\
$k_B$              & $ \bra{|\nab\BBhat|^2}^{1/2}$ & BPS \\
\hline
\end{tabular}
\end{table}

In a recent paper, 
\cite{Kriel+22} determined both $k_\eta^{\rm fit}$
and $k_\nu^{\rm fit}$ and found their ratio to obey
\begin{equation}
k_\eta^{\rm fit}/k_\nu^{\rm fit}=C\,\Pm^{1/2},
\end{equation}
where $C\approx0.88\pm0.23$; see their Eq.~(16).
Using our scaling relations \Eqs{ketaknu}{Params}, we find
\begin{equation}
k_\eta^{\rm fit}\approx 0.093\,k_\nu\,(\Pm/\Pm^{\rm crit})^{1/2}
\approx0.17\,k_\nu\,\Pm^{1/2},
\end{equation}
and, using \Eq{knufit}, we have $C=0.17/0.5=0.34$, which is smaller than
their value.

\subsection{Prandtl number dependence in the nonlinear regime}
\label{NonlinPrandtl}

\begin{table*}
\centering
\caption{
Summary of the nonlinearly saturated runs presented in this paper.
}\label{TsumN}
\begin{tabular}{ccrrrccccccr}
\hline
$\!\!$Run$\!\!$ & $\Ma$ & $\Rey_\lambda$ & $\Rey$ & $\Rm$ & $\Pm$ & $\tilde{k}_\nu$ & $\tilde{k}_B$ &
$\tilde{k}_\eta$ & $\tilde{k}_\eta^{\rm NL}$ & $\Delta t/\tau$ & $N$ \\
\hline
a & 0.068 &   9 &    9 &  900 & 100  &$5.2\pm0.2$& $70\pm3$ &$  95\pm10$ & $150\pm2$ &  60 & 512 \\
b & 0.075 &  19 &   24 &  960 & 40   &$9.8\pm0.2$& $75\pm2$ &$  95\pm10$ & $157\pm2$ &  70 & 512 \\
c & 0.080 &  32 &   52 & 1100 & 20   &$  16\pm1 $& $77\pm2$ &$ 100\pm10$ & $162\pm1$ & 106 & 512 \\
d & 0.087 &  57 &  110 & 1100 & 10   &$  26\pm1 $& $80\pm1$ &$ 105\pm10$ & $163\pm4$ & 106 & 512 \\
e & 0.093 & 110 &  300 & 1200 &  4   &$  51\pm1 $& $80\pm2$ &$ 120\pm10$ & $165\pm2$ & 116 & 512 \\
f & 0.095 & 170 &  615 & 1230 &  2   &$ 113\pm3 $& $83\pm1$ &$ 125\pm5$  & $167\pm2$ & 643 & 512 \\
g & 0.095 & 260 & 1230 & 1230 &  1   &$ 132\pm3 $& $77\pm3$ &$ 132\pm5$  & $168\pm3$ & 191 & 512 \\
g'& 0.101 & 270 & 1310 & 1310 &  1   &$ 139\pm3 $& $78\pm4$ &$ 139\pm5$  & $174\pm2$ & 185 &1024 \\
h & 0.103 & 330 & 1340 &  670 & 0.5  &$ 128\pm1 $& $59\pm4$ &$ 105\pm5$  & $103\pm1$ & 700 & 512 \\
j & 0.111 & 400 & 1440 &  360 & 0.25 &$ 126\pm3 $& $47\pm2$ &$  85\pm5$  &  $62\pm1$ & 493 & 512 \\
l & 0.125 & 460 & 4000 &  400 & 0.1  &$ 329\pm24$& $89\pm6$ &$ 210\pm30$ &  $46\pm2$ &  33 &1024 \\
\hline
\end{tabular}
\end{table*}

\begin{figure}\begin{center}
\includegraphics[width=\columnwidth]{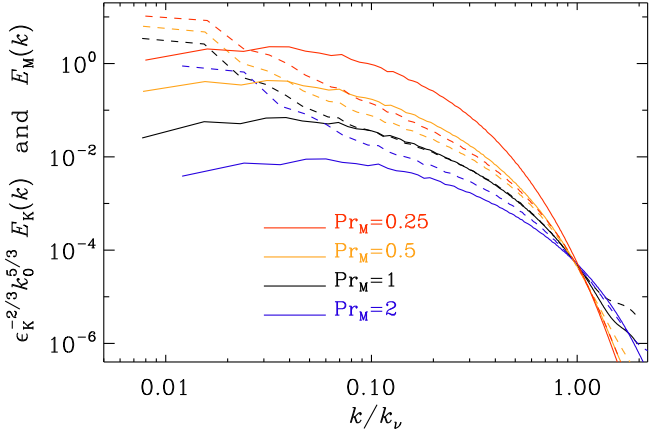}
\end{center}\caption[]{
Magnetic energy spectra (solid lines) for nonlinearly saturated runs
with $\Pm=0.25$ (Run~K, red),
0.5 (Run~J, orange), 1 (Run~I, black), and 2 (Run~H, blue) along
with the corresponding kinetic energy spectra (dashed lines).
}\label{pspec_comp_PrMcrit_sat}\end{figure}

We repeat a sequence of runs similar to that presented in
\Fig{pspec_comp_PrMcrit}, where we show that the shapes of
$\EK(k)$ and $\EM(k)$ 
are similar for 
$\Pm=\Pm^{\rm crit}\approx0.3$.
In the nonlinear regime, the situation is more complicated in
that now also $\EK(k)$ changes near saturation.
The nonlinear runs are denoted analogously to the kinematic
case using, however, lowercase letters.
They are summarized in \Tab{TsumN}, where all data are averaged over a
statistically steady interval of length $\Delta t$.
Here we also define a magnetic dissipation wavenumber analogous to
\Eq{knudef}, i.e.,
\begin{equation}
k_\eta^{\rm NL}=\left(\epsM/\rho_0\eta^3\right)^{1/4},
\label{ketadef}
\end{equation}
where the superscript NL should remind us that this quantity can only be
defined in the nonlinear regime (because otherwise $\epsM\to0$) and that
$k_\eta^{\rm NL}$ is different from the $k_\eta$ defined by collapsing
the curves $\EM(k/k_\eta)$ on top of each other, as in \Fig{pspec_comp2}.
In \Fig{pspec_comp_PrMcrit_sat}, we compare the shapes of $\EM(k)$
and $\EK(k)$ after having scaled them such that their values agree
near $k=k_\nu$.
This scaling allows us to see more readily the relative change of slopes
between $\EM(k)$ and $\EK(k)$.
We see that their profiles now agree with each other for $\Pm=1$.
For larger values of $\Pm$, the slope of the magnetic spectrum is
smaller than 
that of the kinetic energy spectrum, 
while for smaller values
of $\Pm$, the magnetic slopes are steeper.
From this, we conclude that there is a critical value of the magnetic Prandtl number in the nonlinear regime
that is of the order of unity.
This result agrees with that of \cite{Kriel+22}.

\begin{figure}\begin{center}
\includegraphics[width=\columnwidth]{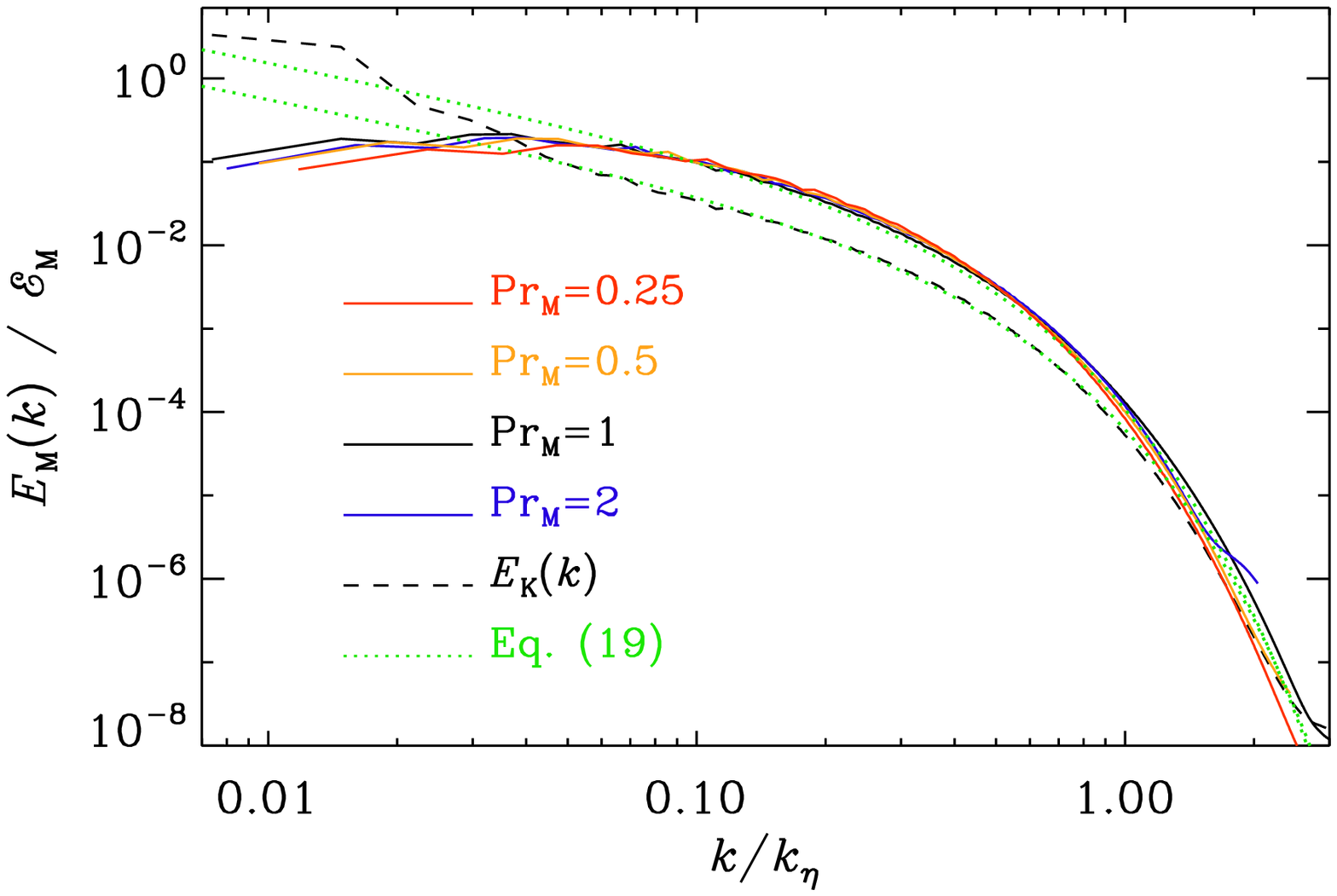}
\end{center}\caption[]{
Similar to \Fig{pspec_comp2}, but for the nonlinearly saturated case.
}\label{pspec_comp2_sat}\end{figure}
The two green lines represent the fits with \Eq{EKfitNL} using the
parameters in \Eq{kNLrel}.
In \Fig{pspec_comp2_sat}, we show the results of collapsing the
nonlinearly saturated spectra on top of each other.
Here, we use Run~g with $\Pm=1$ as reference run,
because this
value is close to the nonlinear value of $\Pm^{\rm crit}$.
The resulting values of $k_\eta$ are listed in \Tab{TsumN}.
In this nonlinear case, \Eq{DottedLineFit} no longer provides
a useful description of the magnetic energy spectrum.
Instead, the dissipative subrange can well be described by a formula
similar to \Eq{EKfitold}, but with a $k^{-1}$ inertial range, i.e.,
\begin{equation}
\EKM(k)\propto k^{-1}\exp(-k/k_{\nu/\eta}^{\rm NLfit}),
\label{EKfitNL}
\end{equation}
where we find
\begin{equation}
k_\nu^{\rm NLfit}\approx0.22\,k_\eta, \quad
k_\eta^{\rm NLfit}\approx0.20\,k_\eta
\quad\mbox{(for $\Pm=1$).}
\label{kNLrel}
\end{equation}
As in the kinematic case, $k_\eta$ depends on $k_\nu$ and $\Pm$,
as will be discussed next.

\begin{figure}\begin{center}
\includegraphics[width=\columnwidth]{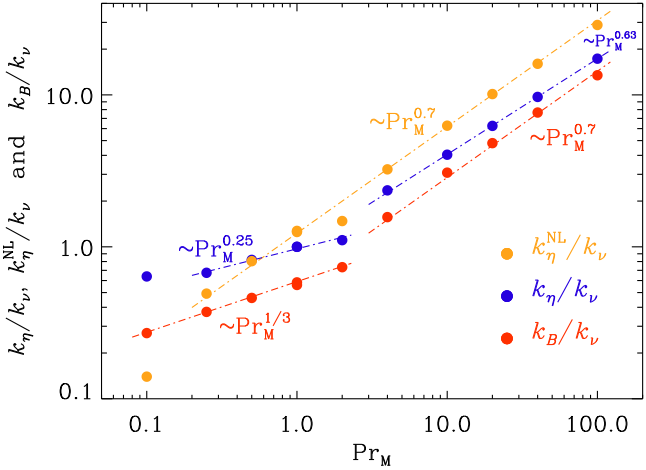}
\end{center}\caption[]{
Similar to \Fig{ptable_keta}, showing
the dependence of $k_\eta/k_\nu$ (closed symbols),
the dependence of $k_\eta^{\rm NL}/k_\nu$ (closed symbols),
and $k_B/k_\nu$ (open symbols) on $\Pm$.
The dashed-dotted lines show $\Pm^{0.3}$ and $\Pm^{0.7}$
scalings for comparison.
}\label{ptable_keta_NL}\end{figure}

In \Fig{ptable_keta_NL}, we plot the dependence of $k_\eta/k_\nu$,
$k_\eta^{\rm NL}/k_\nu$, and $k_B/k_\nu$ on $\Pm$.
We see that now the slopes are different from those in the kinematic case.
Specifically, we find $k_\eta/k_\nu\approx0.95\,\Pm^{0.63}$ for
$\Pm>3$ and $k_\eta/k_\nu\approx0.97\,\Pm^{0.25}$ for $\Pm<3$.
Furthermore, we find $k_\eta^{\rm NL}/k_\nu\approx0.57\,\Pm^{0.7}$ for
$\Pm>3$ and $k_\eta^{\rm NL}/k_\nu\approx0.59\,\Pm^{1/3}$ for $\Pm<3$.
The relations in \Eq{kNLrel} remain approximately valid in the
neighborhood of $\Pm=1$, but deteriorate significantly for $\Pm\gg1$.
This is because the functional form of \Eq{EKfitNL} no longer provides
a good description.
We refer here to earlier work \citep{Bra09,Bra11,Bra14}, where the $\Pm$
dependence in the nonlinear regime has been studied in much more detail.

\section{Conclusions}

In this paper, we have determined the magnetic Prandtl number dependence
for three rather different scales characterizing the dissipative
magnetic structures in a kinematic small-scale dynamo: their diameter,
their theoretical cutoff wavenumbers based on the growth rate, and the
actual spectral cutoff.
For a magnetic Prandtl number of about 0.3, viscous and resistive cutoff
scales are found to be approximately equal.
This is different from the results in the nonlinear regime, where
a critical value of unity is found.
A scaling of the cutoff wavenumber proportional to $\Pm^{1/2}$ is found
for $0.05\leq\Pm\leq100$.
A change of such a scaling is expected for very small values of $\Pm$,
but this cannot be confirmed for moderately small values.

For the actual thickness of flux tubes, we do find a break in the scaling
for $\Pm\approx1$, but it is now steeper than expected both for small
and large values of $\Pm$.
For the scale based on the theoretically expected eigenfunction of the
Kazantsev small-scale dynamo, we also found a slightly steeper scaling,
but no breakpoint for smaller values of $\Pm$ close to $\Pm=0.05$.

For the large values of $\Pm$ that are expected to occur in the
interstellar medium and in galaxy clusters, the viscous scale is much
larger than the resistive one and it may be observationally accessibility
through an excess of the parity-even $E$ polarization over the party-odd
$B$ polarization in synchrotron emission \citep{Bra+22}.
The resistive scale, on the other hand, may be accessible through
interstellar scintillation measurements of pulsars \citep{Cordes+85,
Rickett90, Bhat+04}, as discussed in the introduction.
Thus, there may be ways of comparing theory with observations in the
not too distant future.

It would also be interesting to extend the present study to other
measures of magnetic structures.
One such possibility is the use of Minkowski functionals \citep{Sahni+98}.
\cite{Wilkin+07} have used this method to show that the thickness,
width, and length of magnetic structures from a small-scale dynamo scale
differently with magnetic Reynolds number.
In their case, however, the value of $\Rey$ was held constant, so $\Pm$
and $\Rm$ did not vary independently.
Furthermore, they did not actually solve the momentum equation and
considered instead a prescribed flow with a given power spectrum.
Subsequent work by \cite{Seta+20} demonstrated, however, that
both the thickness and width of the structures show $\Rm^{-1/2}$ scaling.
Furthermore, the structures are more space filling \citep{Seta+Federrath21}.

\section*{Acknowledgements}

We thank James Beattie, Christoph Federrath, Maarit Korpi-Lagg, Neco
Kriel, Joseph Lazio, Matthias Rheinhardt, Alex Schekochihin, Amit Seta,
J\"orn Warnecke, and the anonymous referee for useful discussions and
comments on earlier versions of our work.
This work emerged during discussions at the Nordita program on
``Magnetic field evolution in low density or strongly stratified plasmas''
in May 2022.
The research was supported by the Swedish Research Council
(Vetenskapsr{\aa}det, 2019-04234).
Nordita is sponsored by Nordforsk.
AB and IR would like to thank the Isaac Newton Institute for Mathematical
Sciences, Cambridge, for support and hospitality during the programme
``Frontiers in dynamo theory: from the Earth to the stars'', where the
final version of this paper was completed.
This work was supported by EPSRC grant no EP/R014604/1.34.
We acknowledge the allocation of computing resources provided by the
Swedish National Allocations Committee at the Center for Parallel
Computers at the Royal Institute of Technology in Stockholm and
Link\"oping.
J.S.\ acknowledges the support by the Swiss National Science Foundation
under Grant No.\ 185863.

\section*{Data Availability}

The source code used for the simulations of this study,
the {\sc Pencil Code} \citep{JOSS}, is freely available on
\url{http://github.com/pencil-code/}.
The DOI of the code is http://doi.org/10.5281/zenodo.2315093.
Supplemental Material and the simulation setups with the
corresponding secondary data are available on
\url{http://doi.org/10.5281/zenodo.7090887}; see also
\url{http://www.nordita.org/~brandenb/projects/keta_vs_PrM}
for easier access to the same material as on the Zenodo site
\citep{DATA}.

\bibliographystyle{mnras}
\bibliography{ref}

\newpage
\appendix

\section{Viscous cutoff scaling}
\label{knuScaling}

\begin{figure}\begin{center}
\includegraphics[width=\columnwidth]{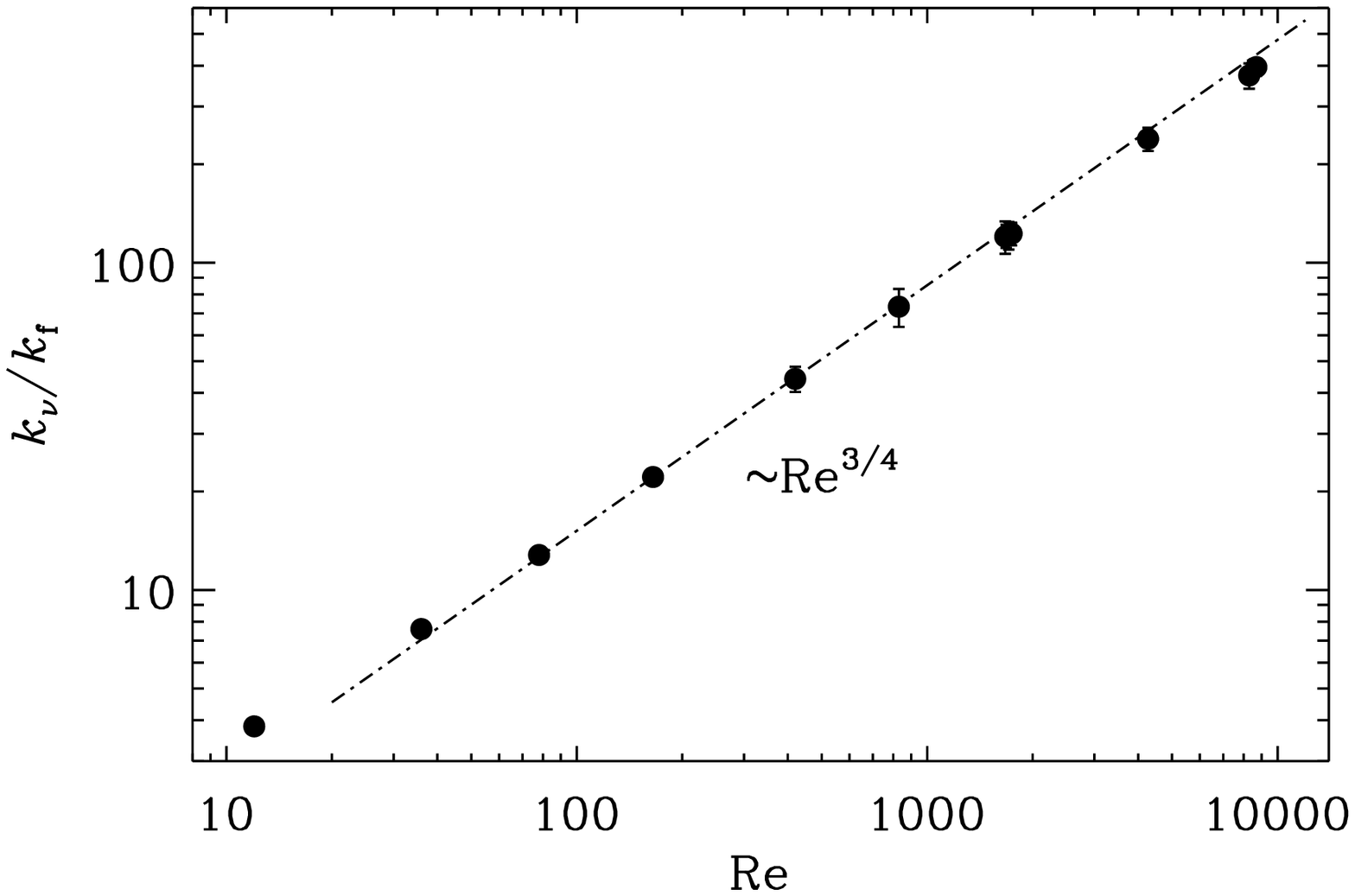}
\end{center}\caption[]{
Dependence of $k_\nu$ on $\Rey$ with $k_\nu/\kf\approx0.48\,\Rey^{3/4}$.
For $\Rey\approx10$, there is no longer a proper turbulent cascade
and the $3/4$ scaling is expected to become invalid.
We have therefore excluded this point from the fit and have only drawn
this point so that one can see that the departure is not very large yet.
The error bars can hardly be noticed.
}\label{ptable_keta_knu}\end{figure}

In \Sec{Spectra} we discussed the expected $\Rey^{3/4}$
scaling of the viscous cutoff wavenumber.
This scaling was also verified by \cite{Kriel+22}, but they did not
actually compute $\epsK$, nor did they use \Eq{knudef}.
To verify that $k_\nu$ obeys this scaling, we show in
\Fig{ptable_keta_knu} the dependence of $k_\nu$ on $\Rey$.
Quantitatively, we have $k_\nu/\kf\approx0.48\,\Rey^{3/4}$.
Small departures are seen for very small and very large values
of $\Rey$.
The latter could be related to insufficient resolution for such a high
value of $\Rey$, while the former could indicate that the 3/4 scaling
is not yet applicable.

Our coefficient in the relation between $k_\nu/\kf$ is larger than that
found by \cite{Kriel+22}.
They found $k_\nu/\kf\approx0.025\,\Rey_{\rm KBSF}^{3/4}$, where
$\Rey_{\rm KBSF}=2\pi\Rey$.
Thus, their relation corresponds to $k_\nu/\kf\approx0.10\,\Rey^{3/4}$.
However, if their effective $\kf$ was also $1.5\,k_1$, as in our case,
then the prefactor would be 0.07 instead of 0.1.

\section{Viscous cutoff wavenumber}
\label{KineticCutoff}

\begin{figure}\begin{center}
\includegraphics[width=\columnwidth]{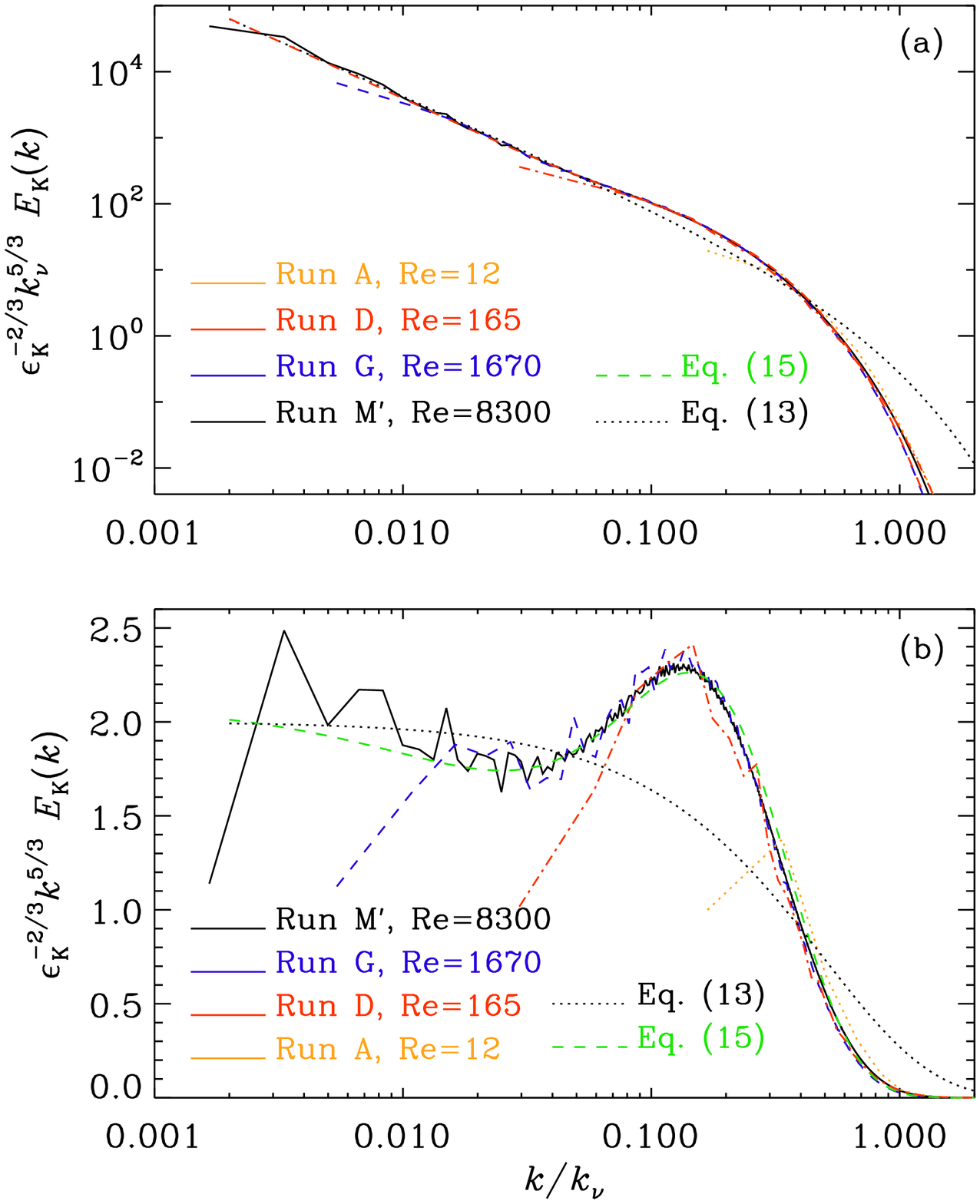}
\end{center}\caption[]{
Kinetic energy spectra for Runs~A, D, G, and M'.
together with the fits given by \Eqs{EKfitold}{EKfit}.
Panel~(b) shows the compensated version of the spectrum shown
in panel~(a).
No adjustable parameters are used, except for the fits
with \Eqs{EKfitold}{EKfit}.
}\label{pspec_comp2_kin}\end{figure}

In \Sec{Redefining}, we discussed different variants of $k_\nu$.
In \Fig{pspec_comp2_kin}(a), we plot kinetic energy spectra for
Runs~A, D, G, and M', together with the fit given by \Eq{EKfitold}
with $k_\nu^{\rm fit}\approx0.5\,k_\nu$ and \Eq{EKfit} with
$n_{\rm bot}=1.8$, $n_{\rm dis}=0.86$,
$k_\nu^{\rm bot}\approx0.056\,k_\nu$, and
$k_\nu^{\rm dis}\approx0.073\,k_\nu$.
The latter corresponds to another definition of the viscous cutoff
wavenumber as $k_\nu^{\rm dis}$.
To demonstrate more clearly the existence of the bottleneck effect in
our simulations, we show in \Fig{pspec_comp2_kin}(b) compensated kinetic
energy spectra, $\epsK^{-2/3} k^{5/3} \EK(k)$, and compare with the fit
given by \Eq{EKfit}.

\bsp	
\label{lastpage}
\end{document}